\documentclass[12pt]{myiopart}
\usepackage{amssymb}
\usepackage{accents} 

\DeclareMathAccent{\wtilde}{\mathord}{largesymbols}{"65}
\DeclareMathAccent{\what}{\mathord}{largesymbols}{"62}

\def\m@th{\mathsurround=0pt}
\mathchardef\bracell="0365 
\def\upbrall{$\m@th\bracell$}
\def\undertilde#1{\mathop{\vtop{\ialign{##\crcr
    $\hfil\displaystyle{#1}\hfil$\crcr
     \noalign
     {\kern1.5pt\nointerlineskip}
     \upbrall\crcr\noalign{\kern1pt
   }}}}\limits}
\def\theequation{\arabic{section}.\arabic{equation}}
\newcommand{\wb}[1]{\overline{#1}}
\newcommand{\ub}[1]{\underline{#1}}
\newcommand{\wh}{\widehat}
\newcommand{\wt}{\widetilde}
\newcommand{\ut}{\undertilde}

\newcommand{\sn}{\mathrm{sn}}
\newcommand{\cQ}{\mathcal{Q}} 
\newcommand{\cH}{\mathcal{H}}

\newcommand{\sss}{\mathfrak{s}}
\newcommand{\ssp}{\mathfrak{p}}
\newcommand{\ssq}{\mathfrak{q}}
\newcommand{\ssr}{\mathfrak{r}}
\newcommand{\sst}{\mathfrak{t}}

\newcommand{\ssl}{\mathfrak{l}}
\newcommand{\sse}{\mathfrak{e}}

\begin{document}
\letter{Seed and soliton solutions for Adler's lattice equation} 
\author{James Atkinson$^1$, Jarmo Hietarinta$^2$ and Frank Nijhoff$^1$}
\address{
$^1$ Department of Applied Mathematics, University of Leeds, Leeds LS2 9JT, UK \\ 
$^2$ Department of Physics, University of Turku, FIN-20014 Turku, FINLAND
}
\begin{abstract}
Adler's lattice equation has acquired the status of a master equation among 2D discrete integrable systems.
In this paper we derive what we believe are the first explicit solutions of this equation.
In particular it turns out necessary to establish a non-trivial seed solution from which soliton solutions can subsequently be constructed using the B\"acklund transformation.
As a corollary we find the corresponding solutions of the Krichever-Novikov equation which is obtained from Adler's equation in a continuum limit.
\end{abstract}

\section{Introduction}
\setcounter{equation}{0} 

In \cite{Adler} V.E. Adler derived a quadrilateral lattice equation as the nonlinear superposition principle 
for B\"acklund transformations (BTs) of the Krichever-Novikov (KN) equation \cite{KN1,KN2}.
This equation can be written as follows, cf.~\cite{Nij}
\begin{equation}
 \eqalign{
A[(u-b)(\wh{u}-b)-(a-b)(c-b)][(\wt{u}-b)(\wh{\wt{u}}-b)-(a-b)(c-b)]+\\ 
B[(u-a)(\wt{u}-a)-(b-a)(c-a)][(\wh{u}-a)(\wh{\wt{u}}-a)-(b-a)(c-a)]\\ 
= ABC(a-b),}
\label{eq:ellform} 
\end{equation}
where $u=u_{n,m},\wt{u}=u_{n+1,m},\wh{u}=u_{n,m+1},\wh{\wt{u}}=u_{n+1,m+1}$ denote the values of a scalar dependent 
variable defined as a function of the independent variables $n,m \in \mathbb{Z}^2$.
The parameters $(a,A)$, $(b,B)$ and $(c,C)$ in (\ref{eq:ellform}) are related points on the Weierstrass elliptic curve 
and can be written in terms of the Weierstrass $\wp$-function:
\begin{equation}
\eqalign{
(a,A)=\left(\wp(\alpha),\wp'(\alpha)\right),\quad 
(b,B)=\left(\wp(\beta),\wp'(\beta)\right),\\ 
(c,C)=\left(\wp(\beta-\alpha),\wp'(\beta-\alpha)\right).
}
\end{equation}
The main integrability property of this equation is that of multidimensional consistency, cf.~\cite{NW,BS}, which 
implies that solutions of eq.~\eref{eq:ellform} can be covariantly embedded in a multidimensional lattice such that 
they obey simultaneously equations of a similar form (albeit with different choices of the lattice parameters) in all 
twodimensional sublattices. Eq.~\eref{eq:ellform} emerged as the most general equation in a classification of scalar 
quadrilateral lattice equations integrable in this sense, \cite{ABS}, which includes the previously known cases 
of lattice equations of Korteweg-de Vries type, cf.~\cite{NQC,NC}.
Furthermore, connections have been established between Adler's equation and other so-called elliptic integrable systems,  
\cite{Q4}, in particular the elliptic Toda Lattice of I. Krichever \cite{ETL} and the elliptic Ruijsenaars-Toda lattice \cite{ERTL}.
Among 2D scalar discrete integrable equations, Adler's equation has thus been revealed as remarkably pervasive, 
making its study of singular importance.

One well established procedure to obtain explicit solutions is the application of BTs on known, perhaps elementary, solutions of 
the lattice equation. This is particularly natural here as the multidimensional consistency implies that the BT is inherent in the lattice 
equation itself. However, in order to implement the BT one needs an initial solution: the \textit{seed} solution, and the establishment 
of elementary solutions that qualify as seed solutions turns out already to be a delicate problem in the case of Adler's equation. 
The resolution of this problem, and the subsequent construction of soliton solutions using the BT, is the 
main achievement of this paper.

\section{The Jacobi form of Adler's equation}
\setcounter{equation}{0} 

We will study Adler's equation in a different form from \eref{eq:ellform}, which was found in \cite{Hie}, namely 
\begin{equation}
\label{eq:hieform} 
p(u\wt{u}+\wh{u}\wh{\wt{u}})-q(u\wh{u}+\wt{u}\wh{\wt{u}})-r(u\wh{\wt{u}}+\wt{u}\wh{u})+pqr(1+u\wt{u}\wh{u}\wh{\wt{u}}) = 0\  , 
\end{equation}
where the parameters $p$, $q$ and $r$ are related to eachother and can be expressed in terms of Jacobi elliptic 
functions with modulus $k$, namely by introducing the points 
\begin{equation}
\eqalign{
(p,P)=(\sqrt{k}\ \sn(\alpha;k),\sn'(\alpha;k)),\quad
(q,Q)=(\sqrt{k}\ \sn(\beta ;k),\sn'(\beta ;k)),\\
(r,R)=(\sqrt{k}\ \sn(\gamma;k),\sn'(\gamma;k)),\quad \gamma=\alpha-\beta\  . 
}
\label{eq:param}
\end{equation}
(the primes denoting derivatives w.r.t.~the first arguments of the Jacobi functions) on the elliptic curve 
\begin{equation}
\Gamma = \left\{(x,X):X^2=x^4+1-(k+1/k)x^2\right\}\  . 
\label{eq:Gamma}
\end{equation}
It was pointed out by V. Adler and Yu. Suris to one of us that the {\it Weierstrass form} (\ref{eq:ellform}) and 
the {\it Jacobi form} (\ref{eq:hieform}) of Adler's equation are equivalent in the sense that there exists a M\"obius 
transformation of the variables together with a bi-rational transformation of the parameters that takes one form into 
the other, \cite{AdlSur}. 

We adopt here the Jacobi form because the analysis leading to the solutions is simpler than in the Weierstrass form. 
For that purpose it is useful 
to view the relation between the parameters in terms of the Abelian group structure on the elliptic curve \eref{eq:Gamma} which can  
be defined through the following rational representation:
\begin{equation}\label{eq:prod} 
\ssp = \ssq*\ssr = \left(\ \frac{qR+rQ}{1-q^2r^2} , \ \frac{Qq(r^4-1)-Rr(q^4-1)}{(1-q^2r^2)(rQ-qR)} \ \right)
\end{equation}
where $\ssp=(p,P), \ssq=(q,Q)$ and $\ssr=(r,R)$. It can be verified by direct computation that the product in the group 
$*$ defined in this way is associative
 and it is obviously commutative, and furthermore the identity is $\sse=(0,1)$ and inverse of a point $\ssp$ is 
$\ssp^{-1}=(-p,P)$. Thus, we consider the lattice parameters of \eref{eq:hieform} to be the points $\ssp=(p,P)$ and 
$\ssq=(q,Q)$ on $\Gamma$, while $\ssr=\ssp*\ssq^{-1}$ which expresses the relation $\gamma=\alpha-\beta$ in terms of the uniformising variables 
introduced in \eref{eq:param}. In fact, this relation encodes the addition formulae for the relevant Jacobi functions, where 
$\ssr$ can be seen as the point obtained by a shift of $\ssp$ on the elliptic curve defined by the point $\ssq$.  
Alternatively, it is well known that a symmetric biquadratic equation defines a shift on an elliptic curve, see for example \cite{BV}, and 
this can be made explicit by introducing the symmetric biquadratic
\begin{equation}
\cH_\ssr(p,q) = \frac{1}{2r}\left( p^2+q^2-(1+p^2q^2)r^2-2pqR \right)\  .
\label{eq:biquad}
\end{equation}
{}From the factorisation 
\begin{equation}\label{eq:Hbiquad}
\left(p-\frac{qR+rQ}{1-q^2r^2}\right)\left(p-\frac{qR-rQ}{1-q^2r^2}\right) = \frac{2r}{1-q^2r^2}\cH_\ssr(p,q).
\end{equation}
we see that $\cH_\ssr(p,q)=0$ is satisfied if $\ssp=\ssq*\ssr$ or $\ssp=\ssq*\ssr^{-1}$, i.e., if $\alpha=\beta\pm\gamma$ in terms of the 
uniformising variables, and in this way $\cH_\ssr=0$ defines an addition formula for Jacobi elliptic functions. The biquadratic 
$\cH_\ssr$ is connected 
to a symmetric \textit{tri}quadratic in terms of the three variables $p$, $q$, $r$, now appearing on equal footing, given by  
\begin{equation}\label{eq:triquad} 
\fl  
H(p,q,r) = \frac{1}{2\sqrt{k}}(p^2\!+q^2\!+r^2\!+p^2q^2r^2)-\frac{\sqrt{k}}{2}(1+p^2q^2\!+q^2r^2\!+r^2p^2)+(k-\frac{1}{k})pqr
\end{equation}
which also defines a shift on the same curve by the statement that $H(p,q,r)=0$ if $\ssp=\ssq*\ssr*\sss^{-1}$ or $\ssp=\ssq*\ssr^{-1}*\sss$ where 
$\sss=(\sqrt{k},0)$, (the latter being a branch point of the curve $\Gamma$). This triquadratic is M\"obius related to a similar triquadratic 
expression for Weierstrass elliptic functions, and the latter, which was also used by Adler in \cite{Adler}, plays a similar role in the analysis 
of \eref{eq:ellform} as the biquadratic $\cH_\ssr$ in the construction for \eref{eq:hieform} that will follow below.  

In terms of the points $\ssp$, $\ssq$ on the curve $\Gamma$ it is now useful to introduce the quadrilateral expression 
\begin{equation}
\fl 
\cQ_{\ssp,\ssq}(u,\wt{u},\wh{u},\wh{\wt{u}}) =  p(u\wt{u}+\wh{u}\wh{\wt{u}})-q(u\wh{u}+\wt{u}\wh{\wt{u}})
-\frac{pQ-qP}{1-p^2q^2}(u\wh{\wt{u}}+\wt{u}\wh{u}-pq(1+u\wt{u}\wh{u}\wh{\wt{u}}))  
\label{eq:convenient}
\end{equation}
in terms of which \eref{eq:hieform} takes the form 
\begin{equation}
\cQ_{\ssp,\ssq}(u,\wt{u},\wh{u},\wh{\wt{u}}) = 0\  . 
\label{eq:generic}
\end{equation}
We note that $\cQ$ and $\cH$ are related by the equations
\begin{equation}
\eqalign{
\cQ_{\ssp,\ssq}(u,\wt{u},\wh{u},\wh{\wt{u}})\cQ_{\ssp,\ssq^{-1}}(u,\wt{u},\wh{u},\wh{\wt{u}})\\
= \frac{4p^2q^2}{1\!-\!p^2q^2}\left(\cH_\ssp(u,\wt{u})\cH_\ssp(\wh{u},\wh{\wt{u}})-\cH_\ssq(u,\wh{u})\cH_\ssq(\wt{u},\wh{\wt{u}})\right),
}
\end{equation}
and
\begin{equation}\label{eq:Qdiscr} 
\cQ_{\wh{u}}\cQ_{\wh{\wt{u}}}-\cQ \cQ_{\wh{u}\wh{\wt{u}}} = 2pqr\cH_\ssp(u,\wt{u})\  , 
\end{equation}
cf.~\cite{ABS}, where in \eref{eq:Qdiscr} we have suppressed the arguments and subscripts of 
$\cQ_{\ssp,\ssq}(u,\wt{u},\wh{u},\wh{\wt{u}})$ in order 
to make room for partial derivatives with respect to the arguments.

\section{Seed solution}
\setcounter{equation}{0} 

The cubic consistency of Adler's equation, \cite{NW,BS}, means that given one solution, $u$, of \eref{eq:generic}
the pair of ordinary difference equations for $v$ 
\begin{equation}
\cQ_{\ssp,\ssl}(u,\wt{u},v,\wt{v}) = 0, \quad \cQ_{\ssq,\ssl}(u,\wh{u},v,\wh{v}) = 0, 
\label{eq:genericBT}
\end{equation}
are compatible and, moreover, if $v$ satisfies this system it is also a solution of \eref{eq:generic}.
We say that \eref{eq:genericBT} forms an auto-B\"acklund transformation (BT) for \eref{eq:generic}.
The new solution $v$ may depend not only on $u$ but also on the B\"acklund parameter $\ssl$ and on one integration constant.

For Adler's equation in Jacobi form \eref{eq:hieform} we can easily verify the solution
\begin{equation}
u=\sqrt{k}\,\sn(\xi_0+n\alpha + m\beta;k), 
\label{eq:badseed} 
\end{equation}
where $\xi_0$ is an arbitrary constant and $\alpha$,$\beta$ are as defined in \eref{eq:param}.
It turns out, however, that this solution is not suitable as a {\it seed} for constructing soliton solutions using the BT.
In fact for \eref{eq:badseed} the equations \eref{eq:genericBT} for $v$ are reducible and have only two solutions which we label $\wb{u}$ and $\ub{u}$, given by
\begin{equation}
\eqalign{
\wb{u}=\sqrt{k}\,\sn(\xi_0+n\alpha+m\beta+\lambda;k),\\
\ub{u}=\sqrt{k}\,\sn(\xi_0+n\alpha+m\beta-\lambda;k),
}
\label{eq:badseedext}
\end{equation}
where $\lambda$ is the uniformizing variable associated with $\ssl$, i.e., $\ssl=(l,L)=(\sqrt{k} \ \sn(\lambda;k),\sn'(\lambda;k))$.
In fact \eref{eq:badseedext} is nothing more than \eref{eq:badseed} extended into a third lattice direction associated with the BT \eref{eq:genericBT}, 
(hence the notation with $\wb{u}$ interpreted as forward shift on $u$ in the direction of the BT, and $\ub{u}$ as a backward shift).
This is an idea which we will use later and refer to as a covariant extension of the solution.
Since the action of the BT only trivially extends \eref{eq:badseed}, application of the BT on the solution 
\eref{eq:badseed} clearly does not lead to essentially new solutions.
Therefore, in order to find soliton solutions for Adler's equation we require a {\it germinating} seed solution, i.e., a solution on which the action of 
the BT is non-trivial.

We proceed by making the hypothesis that there exists a solution of \eref{eq:generic} which is a fixed-point of its BT \eref{eq:genericBT}, i.e., which is 
constant in a third lattice direction associated with a particular B\"acklund parameter $\sst=(t,T)=(\sqrt{k} \ \sn(\theta;k),\sn'(\theta;k))$.
In that case the BT reduces to the system
\begin{equation}
\cQ_{\ssp,\sst}(u,\wt{u},u,\wt{u}) = 0, \quad \cQ_{\ssq,\sst}(u,\wh{u},u,\wh{u}) = 0. 
\label{eq:fixedBT}
\end{equation}
Although cubic consistency by itself does not guarantee that such fixed points exists, i.e., that the system \eref{eq:fixedBT} is compatible, we will show that for 
Adler's equation \eref{eq:hieform} such a solution indeed arises and provides a germinating seed. 

To find such solutions we begin by considering first the equation of \eref{eq:fixedBT} involving the shift $u\mapsto\wt{u}$. From the expression 
\eref{eq:convenient} this is quadratic and symmetric in $u$ and $\wt{u}$, and hence can be viewed as a quadratic correspondence having $n+1$ images on the $n^{\mathrm{th}}$ 
iteration. What is not obvious is that as a multivalued map this correspondence commutes with its counterpart, 
the map $u \mapsto \wh{u}$ defined by the second equation of \eref{eq:fixedBT}, 
but this will become apparent below. Directly from \eref{eq:convenient} we obtain the biquadratic 
\begin{equation}
\fl\qquad 
\cQ_{\ssp,\sst}(u,\wt{u},u,\wt{u}) =  2pu\wt{u}-t(u^2+\wt{u}^2)-\frac{pT-tP}{1-p^2t^2}(2u\wt{u}-pt(1+u^2\wt{u}^2))=0\  ,
\label{eq:Jbiquad}
\end{equation}
and comparing this with the general expression for $\cH_\ssr$, with $p$ replaced by $u$ and $q$ replaced by $\wt{u}$ we 
are led to introduce a new parameter $\ssp_\theta=(p_\theta,P_\theta)$ defined by
\begin{equation}
p_\theta^2 = p \frac{pT-tP}{1-p^2t^2}, \quad P_\theta = \frac{1}{t}\left(p-\frac{pT-tP}{1-p^2t^2}\right)\  .
\label{eq:tp}
\end{equation}
For later convenience we use the symbolic notation $\delta_\theta(\ssp,\ssp_\theta)$ to denote the correspondence between the 
original parameter $\ssp$ and the $\theta$-deformed parameter $\ssp_\theta$, given by the relations \eref{eq:tp}. 
Thus, we can now write 
\begin{equation}
\fl\quad 
\cQ_{\ssp,\sst}(u,\wt{u},u,\wt{u}) = t\left(2u\wt{u}P_\theta - u^2 - \wt{u}^2 + (1+u^2\wt{u}^2)p_\theta^2\right) = -2p_\theta t \cH_{\ssp_\theta}(u,\wt{u})=0
\label{eq:Jbiquad2}
\end{equation}
bringing \eref{eq:Jbiquad} to the standard form of the biquadratic addition formula \eref{eq:biquad} for the Jacobi elliptic functions. 
As before, this defines a group law on an elliptic curve of the same type as \eref{eq:Gamma} but with now a new modulus $k_\theta$, namely 
\begin{equation}
\eqalign{
\Gamma_\theta = \left\{ (x,X) : X^2 = x^4 + 1 - (k_\theta+1/k_\theta)x^2 \right\},\\
k_\theta+\frac{1}{k_\theta} = 2\frac{1-T}{t^2} = 2\frac{1-\sn'(\theta;k)}{k\,\sn^2(\theta;k)}\  .
}
\label{eq:seed_curve}
\end{equation}
The deformed parameter $\ssp_\theta$ is a point of this deformed curve, $\ssp_\theta\in\Gamma_\theta$, as can be verified  by direct computation using the fact that 
the seed parameter and the original lattice parameter lie on the original curve, ~$\sst,\ssp\in\Gamma$~. What 
\eref{eq:Jbiquad2} tells us, is that the solution 
of the first part of the fixed point equation can be parametrised in terms of Jacobi elliptic funtions associated with the deformed curve $\Gamma_\theta$, and thus 
by introducing uniformizing variables on $\Gamma_\theta$ we can write the solution explicitly as follows.  
Setting $\ssp_\theta = (\sqrt{k_\theta} \ \sn(\alpha_\theta;k_\theta),\sn'(\alpha_\theta;k_\theta))$, $u = \sqrt{k_\theta} \ \sn(\xi_\theta;k_\theta)$ and then
\begin{equation}
\wt{\xi}_\theta = \xi_\theta \pm \alpha_\theta \ \ \Longrightarrow \ \ \cH_{\ssp_\theta}(u,\wt{u})=0 \ \ \Longrightarrow \ \ \cQ_{\ssp,\sst}(u,\wt{u},u,\wt{u})=0. 
\label{eq:seed1}
\end{equation}  
Here the deformed lattice parameter $\alpha_\theta$ is related in a transcendental way to the original lattice 
parameter $\alpha$ through either of the equivalent relations:
\begin{equation}\label{eq:defparm}
\fl\quad 
k_\theta\,\sn^2(\alpha_\theta;k_\theta)=k\,\sn(\alpha;k)\sn(\alpha-\theta;k)\  ,\quad
\sn'(\alpha_\theta;k_\theta)=\frac{\sn(\alpha;k)-\sn(\alpha-\theta;k)}{\sn(\theta;k)}\  . 
\end{equation} 

However, so far we have only dealt with the first part of \eref{eq:fixedBT}; we must also solve simultaneously the second part. The crucial observation is that 
deformed curve that emerges in the solution of the first part is \textit{independent of the lattice parameter $\ssp$ characterising the direction in the lattice}. 
Thus, if we proceed in precisely the same way with solving the second equation in \eref{eq:fixedBT} we get exactly the same curve $\Gamma_\theta$ and hence 
the parametrisation of the solutions of the latter can are given by the shifts on the curve as follows 
\begin{equation}
\wh{\xi}_\theta = \xi_\theta \pm \beta_\theta \ \ \Longrightarrow \ \ \cH_{\ssq_\theta}(u,\wh{u})=0 \ \ \Longrightarrow \ \ \cQ_{\ssq,\sst}(u,\wh{u},u,\wh{u})=0.
\label{eq:seed2}
\end{equation}
where we have now introduced the deformed variable $\ssq_\theta$, again through the correspondence $\delta_\theta(\ssq,\ssq_\theta)$, and 
parametrizing $\ssq_\theta = (\sqrt{k_\theta} \ \sn(\beta_\theta;k_\theta), \sn'(\beta_\theta;k_\theta))$ in terms of 
a deformed uniformizing variable $\beta_\theta$. The latter parameter obeys similar relations to \eref{eq:defparm} with $\alpha$ 
replaced by $\beta$, but involving the same modulus $k_\theta$. 

Now, clearly on the curve $\Gamma_\theta$ the maps given in \eref{eq:seed1} and \eref{eq:seed2} commute, and this implies the compatibility of the 
solutions of both members of \eref{eq:fixedBT}, i.e., we have a simultaneous solution which is the required seed solution we are looking for. 
In explicit form this new seed solution can be expressed as 
\begin{equation}
u_\theta(n,m) = \sqrt{k_\theta} \ \sn(\xi_\theta;k_\theta), 
\quad \xi_\theta = \xi_{\theta,0} +n\alpha_\theta + m\beta_\theta\  ,
\label{eq:seed}
\end{equation}
where $\xi_{\theta,0}$ is an arbitrary integration constant. 
We note that the seed solution $u_\theta$, distinguished by the label $\theta$, reduces to the non-germinating seed 
\eref{eq:badseed} in the limit $\theta\longrightarrow 0$, i.e., $\sst\longrightarrow\sse=(0,1)$, the unit on the curve.
The equations \eref{eq:seed1} and \eref{eq:seed2} clearly have other solutions as well, corresponding to the choice of 
sign at each iteration, and these lead to different seeds. As a particular example, an \textit{alternating seed solution} 
can be obtained by chosing a flip of sign at each iteration step of the seed map.    
We will refer to \eref{eq:seed} as the canonical seed solution.

\section{One-soliton solution}
\setcounter{equation}{0} 

We will now show that the canonical seed solution germinates by applying the BT to it, i.e., by computing the one-soliton solution.
We need to solve the simultaneous ordinary difference equations in $v$
\begin{equation}
\cQ_{\ssp,\ssl}(u_\theta,\wt{u}_\theta,v,\wt{v}) = 0, \quad \cQ_{\ssq,\ssl}(u_\theta,\wh{u}_\theta,v,\wh{v}) = 0, 
\label{eq:solitonBT}
\end{equation}
which define the BT $u_\theta \mapsto v$ with B\"acklund parameter $\ssl$.
The seed itself can be covariantly extended in the lattice direction associated with this BT, that is we may complement \eref{eq:seed} with the equation $\wb{\xi}_\theta=\xi_\theta+\lambda_\theta$, where the $\wb{\phantom{u}}$ denotes a shift in this new direction which is associated with the parameter $\ssl$ and as before $\lambda_\theta$ is the uniformizing variable for $l_\theta$ defined by the relation $\delta_\theta(\ssl,\ssl_\theta)$.
The problem of solving the system \eref{eq:solitonBT} can be simplified because this covariantly extended seed provides two particular solutions, i.e.,
\begin{equation}
\eqalign{
\cQ_{\ssp,\ssl}(u_\theta,\wt{u}_\theta,\wb{u}_\theta,\wt{\wb{u}}_\theta) = 0, \quad
\cQ_{\ssq,\ssl}(u_\theta,\wh{u}_\theta,\wb{u}_\theta,\wh{\wb{u}}_\theta) = 0, \\ 
\cQ_{\ssp,\ssl}(u_\theta,\wt{u}_\theta,\ub{u}_\theta,\wt{\ub{u}}_\theta) = 0, \quad
\cQ_{\ssq,\ssl}(u_\theta,\wh{u}_\theta,\ub{u}_\theta,\wh{\ub{u}}_\theta) = 0.}
\label{eq:partsol}
\end{equation}
(Compare with the non-germinating seed for which these were the only solutions.)
From the multilinearity of \eref{eq:convenient} the equations \eref{eq:solitonBT} are discrete Riccati equations for $v$.
The key observation is that since these equations share two solutions \eref{eq:partsol} they can be simultaneously reduced to homogeneous linear equations for a new variable $\rho$ by the substitution
\begin{equation}
v = \frac{1}{1-\rho}\ \ub{u}_\theta-\frac{\rho}{1-\rho}\ \wb{u}_\theta.
\label{eq:vsub}
\end{equation}
After some manipulation the system for $\rho$ found by substituting \eref{eq:vsub} into \eref{eq:solitonBT} can be written as 
\begin{equation}
\fl\qquad 
\wt{\rho} = \Bigg(\frac{p_\theta l - l_\theta p}{p_\theta l + l_\theta p}\Bigg)\Bigg(\frac{1-l_\theta \wb{p}_\theta u_\theta \wt{u}_\theta}{1+l_\theta \ub{p}_\theta u_\theta \wt{u}_\theta}\Bigg)\rho\quad , \quad
\wh{\rho} = \Bigg(\frac{q_\theta l - l_\theta q}{q_\theta l + l_\theta q}\Bigg)\Bigg(\frac{1-l_\theta \wb{q}_\theta u_\theta \wh{u}_\theta}{1+l_\theta \ub{q}_\theta u_\theta \wh{u}_\theta}\Bigg)\rho,
\label{eq:plane-waves}
\end{equation}
where we mildly abuse notation by introducing the modified parameters
\begin{equation}
\eqalign{
\wb{p}_\theta=\sqrt{k_\theta} \ \sn(\alpha_\theta+\lambda_\theta;k_\theta),\quad
\ub{p}_\theta=\sqrt{k_\theta} \ \sn(\alpha_\theta-\lambda_\theta;k_\theta),\\
\wb{q}_\theta=\sqrt{k_\theta} \ \sn(\beta_\theta+\lambda_\theta;k_\theta),\quad
\ub{q}_\theta=\sqrt{k_\theta} \ \sn(\beta_\theta-\lambda_\theta;k_\theta)
}
\end{equation}
($p_\theta$ and $q_\theta$ do not depend on lattice shifts).
We take \eref{eq:plane-waves} as the defining equations for $\rho$, which we refer to as the plane-wave factor.
The compatibility of this system for $\rho$ can be verified directly, specifically $\wt{\wh{\rho}}=\wh{\wt{\rho}}$ as a consequence of the identity for the Jacobi $\sn$ function
\begin{equation*}
\eqalign{
\left(\frac{\scriptstyle{1-k^2\sn(\lambda)\sn(\alpha+\lambda)\sn(\xi)\sn(\xi+\alpha)}}
           {\scriptstyle{1+k^2\sn(\lambda)\sn(\alpha-\lambda)\sn(\xi)\sn(\xi+\alpha)}}\right)
\left(\frac{\scriptstyle{1-k^2\sn(\lambda)\sn(\beta+\lambda)\sn(\xi+\alpha)\sn(\xi+\alpha+\beta)}}
           {\scriptstyle{1+k^2\sn(\lambda)\sn(\beta-\lambda)\sn(\xi+\alpha)\sn(\xi+\alpha+\beta)}}\right) =  \\
\left(\frac{\scriptstyle{1-k^2\sn(\lambda)\sn(\beta+\lambda)\sn(\xi)\sn(\xi+\beta)}}
           {\scriptstyle{1+k^2\sn(\lambda)\sn(\beta-\lambda)\sn(\xi)\sn(\xi+\beta)}}\right) 
\left(\frac{\scriptstyle{1-k^2\sn(\lambda)\sn(\alpha+\lambda)\sn(\xi+\beta)\sn(\xi+\alpha+\beta)}}
           {\scriptstyle{1+k^2\sn(\lambda)\sn(\alpha-\lambda)\sn(\xi+\beta)\sn(\xi+\alpha+\beta)}}\right). 
}
\end{equation*}
The one-soliton solution for the Jacobi form of Adler's equation, which we denote by $u^{(1)}$, is thus given by
\begin{equation}
u^{(1)}(n,m) = \frac{\sqrt{k_\theta}}{1-\rho}\left( \sn(\xi_\theta-\lambda_\theta;k_\theta) - \rho \ \sn(\xi_\theta+\lambda_\theta;k_\theta) \right)\ ,
\label{eq:soliton}
\end{equation}
with $\xi_\theta$ as in \eref{eq:seed} and $\rho$ defined by \eref{eq:plane-waves}.

\section{Compatible continuous systems}
\setcounter{equation}{0} 

As was pointed out in \cite{ABS} Adler's equation goes to the Krichever-Novikov (KN) equation in a particular continuum limit.
We give this limit by first introducing the formal analytic difference operator
\begin{equation}
C_p = e^{\sqrt{2p}(\partial_x+\frac{p}{6}\partial_y)}
\end{equation}
and hence continuous variables $x$ and $y$.
By writing $\wt{u}=C_p u$ and $\wh{u}=C_q u$ the variable $u$ on the lattice can be reinterpreted as a sampling of $u$ at points on the $(x,y)$ plane.
In the limit as $\ssp, \ssq \longrightarrow (0,1),(0,1)$ we find
\begin{eqnarray}
\cQ_{\ssp,\ssq}(u,C_pu,C_qu,C_pC_qu) = 0 \ \longrightarrow \nonumber \\
u_y-u_{xxx}+\frac{3}{2 u_x}\left(u_{xx}^2-u^4-1+\left(k+\frac{1}{k}\right)u^2\right) = 0, \label{eq:KN}
\end{eqnarray}
i.e., Adler's equation goes to the KN equation.
In the same formal limit the equations \eref{eq:genericBT} and \eref{eq:fixedBT} go to
\begin{eqnarray}
&u_x v_x - \cH_\ssl(u,v) = 0 \ \ \textrm{and}& \label{eq:KNBT}\\
&u_x^2 - \cH_\sst(u,u) = 0& \label{eq:KNseedeq}
\end{eqnarray}
respectively, that is the auto-BT for the KN equation, cf.~\cite{Adler}, and the equation for its seed solution.
We note that other compatible differential-difference equations can also be obtained in this way, for example by writing 
$\wh{u}=C_qC_p^{-1}\wt{u}$ and taking the limit $\ssq\longrightarrow \ssp$, Adler's equation \eref{eq:hieform} goes to 
\begin{equation}
p(\wt{u}-\ut{u})u_z - P(\wt{u}+\ut{u})u + \wt{u}\ut{u} + u^2 - p^2(1+\wt{u}\ut{u}u^2) = 0
\end{equation}
where $z = \sqrt{2p}(x+2y/p)/P$.

To compute the seed solution of \eref{eq:KN} one can solve \eref{eq:KNseedeq} coupled with \eref{eq:KN} itself, or alternatively 
one can take a continuum limit of the seed solution for Adler's equation \eref{eq:seed}.
Either way the calculation is straightforward and we find
\begin{equation}
\eqalign{
u_\theta(x,y) = \sqrt{k_\theta} \ \sn\left(\xi_\theta(x,y);k_\theta\right), \\
\xi_\theta(x,y) = \sqrt{-t/2}\left(x_0+x-(2+T)y/t\right)/\sqrt{k_\theta}.
}
\label{eq:KNseed}
\end{equation}
where $x_0$ is an arbitrary integration constant.
This can be verified as a solution of \eref{eq:KN} directly (and for fixed $y$ as the general solution of \eref{eq:KNseedeq}). 

Calculation of the one-soliton solution proceed by the continuum limit of \eref{eq:soliton} and \eref{eq:plane-waves}, or by substitution 
of the continuous seed solution \eref{eq:KNseed} into \eref{eq:KNBT} followed by the identification of two particular solutions (this time 
from an extension of the continuous seed solution defined by \eref{eq:KNseed} together with $\wb{\xi}_\theta(x,y)=\xi_\theta(x,y)+\lambda_\theta$).
This calculation gives the one-soliton solution for \eref{eq:KN} as
\begin{eqnarray}
\fl \qquad 
u^{(1)}(x,y) &=& \frac{\sqrt{k_\theta}}{1-\rho(x,y)}\left( \sn(\xi_\theta(x,y)-
\lambda_\theta;k_\theta) - \rho(x,y) \ \sn(\xi_\theta(x,y)+\lambda_\theta;k_\theta) \right) \label{eq:cont-soliton}
\end{eqnarray}
with $\xi_\theta(x,y)$ as in \eref{eq:KNseed}.
Here $\rho(x,y)$ is the continuous plane-wave satisfying the following ODE in terms of $x$
\begin{equation}
\rho_x(x,y) = \frac{-l_\theta}{l\sqrt{-t/2}}\left(\frac{1-l^2u_\theta^2(x,y)}{1-l_\theta^2u_\theta^2(x,y)}\right) \rho(x,y),
\label{eq:cont-plane-wave}
\end{equation}
and where the $y$ dependence can be found by the substitution of \eref{eq:cont-soliton} into \eref{eq:KN} itself. 
Note that in the above the square roots in $\sqrt{k_\theta}$ and $\sqrt{-t/2}$ refer to same choice of branch wherever 
they appear.

\section{Concluding remarks}
\setcounter{equation}{0} 

In this letter we have given solutions to Adler's lattice equation in its Jacobi form.
The seed solution is found as a fixed-point of the auto B\"acklund transformation (BT) for this equation, and application of the BT (with different 
B\"acklund parameter) to the seed solution yields the one-soliton solution.
The construction of the seed solution requires a deformation of the original elliptic curve in terms of 
which the lattice parameters of the equation were given. It seems that there are interesting modular transformations that arise through this construction. 
The one-soliton solution involves some functions which are solutions of a 
consistent set of first order homogeneous, linear, but non-autonomous, equations involving the seed solution. 

In order to calculate higher soliton solutions it suffices to apply the permutability condition of the 
BTs, which is once again implicit in the original lattice equation \eref{eq:generic} introducing parameters 
$\ssl_1$ and $\ssl_2$ and taking for $u$ the seed solution and for $\wt{u}$ and $\wh{u}$ the one-soliton solutions 
with B\"acklund parameters $\ssl_1$ and $\ssl_2$ respectively. This leads to a 2-soliton solution which is 
a rational function of the seed and the soliton solution. These results and other details will be given in a 
separate publication \cite{later}.
With one exception (the Lattice Schwarzian KdV equation) the methodology presented here works for all equations in the classification of 
Adler, Bobenko and Suris \cite{ABS}, and a list of these solutions has been obtained which will be included in \cite{later}. The results presented here form, 
as far as we are aware, the first examples of explicit 
solutions of Adler's equation.

\end{document}